\documentclass[conference]{IEEEtran}
\IEEEoverridecommandlockouts
\usepackage{cite}
\usepackage{amsmath,amssymb,amsfonts}
\usepackage{algorithmic}
\usepackage{graphicx}
\usepackage{textcomp}
\usepackage{xcolor}
\usepackage{booktabs}

\usepackage{subfigure}
\usepackage{hyperref}
\usepackage{makecell}
\usepackage{adjustbox}

\hypersetup{hidelinks}

\def\BibTeX{{\rm B\kern-.05em{\sc i\kern-.025em b}\kern-.08em
    T\kern-.1667em\lower.7ex\hbox{E}\kern-.125emX}}
\begin{document}

\title{Reconfigurable Quantum Internet Service Provider}

\author{
    \IEEEauthorblockN{Zhaohui Yang\IEEEauthorrefmark{1}, Chaohan Cui\IEEEauthorrefmark{2}}
    
    \IEEEauthorblockA{\textit{\IEEEauthorrefmark{1}Department of Electrical and Computer Engineering, University of Arizona, Tucson, Arizona 85721, USA}}

    \IEEEauthorblockA{\textit{\IEEEauthorrefmark{2}J. C. Wyant College of Optical Sciences, University of Arizona, Tucson, Arizona 85721, USA}}

    \IEEEauthorblockA{\{zhy, chaohancui\}@arizona.edu}
}

\maketitle

\begin{abstract}
    Quantum network is a research topic involving fundamental quantum information science, engineering designs, and experimental implementations. Existing research works primarily pertain to theoretical architecture designs and proof-of-concept experimental demonstrations. With the recent developments in engineering quantum systems, the realization of scalable local-area quantum networks has become viable. However, the design and implementation of a quantum network is a holistic task that is way beyond the scope of an abstract design problem. As such, a testbed on which multiple disciplines can verify the design and implementation across a full networking stack has become a necessary infrastructure for the future development of quantum networks. In this work, we demonstrate the concept of quantum internet service provider (QISP), in analogy to the conventional ISP that allows for the sharing of classical information between the network nodes. The QISP is significant for the next-generation quantum networks as it coordinates the production, management, control, and sharing of quantum information across the end-users of a quantum network. We construct a reconfigurable QISP comprising both the quantum hardware and classical control software. Building on the fiber-based quantum-network testbed of the Center for Quantum Networks (CQN) at the University of Arizona (UA), we develop an integrated QISP prototype based on a Platform-as-a-Service (PaaS) architecture, whose classical control software is abstracted and modularized as an open-source QISP framework. To verify and characterize the QISP's performance, we demonstrate multi-channel entanglement distribution and routing among multiple quantum-network nodes with a time-energy entangled-photon source. We further perform field tests of concurrent services for multiple users across the quantum-network testbed. Our experiment demonstrates the robust capabilities of the QISP, laying the foundation for the design and verification of architectures and protocols for future quantum networks.

\end{abstract}



\section{Introduction}

    Quantum networks have attracted increasing interests in the past decades. Different from  classical networks, quantum networks transmit information following quantum mechanical principles to support promising applications in the area of quantum information science (QIS), such as quantum secret sharing \cite{quantum-secret-sharing-1999}, distributed quantum computing \cite{quantum-internet-2018, distributed-quantum-computing-1999}, quantum-enhanced sensing \cite{quantum-enhanced-measurement-2004, RF-photonic-sensor-network-2020}, and quantum teleportation \cite{bouwmeester1997experimental}. While large-scale quantum networks are still under exploration due to the lack of high-quality quantum repeaters and quantum memories for long-distance transmission, many local and metropolitan-area delicated quantum networks have been deployed \cite{RQLAN-2021,garcia2021madrid,li2021building}. As the scale of quantum networks increases, user-cases definitely become plentiful and heterogeneous, thus leading to challenging requirements for network architectures and components. Instead, by abstracting the homogeneity of near-term quantum applications and leveraging architectural design principles from the classical internet service, a reconfigurable quantum network architecture can be reasonably portrayed. 
    
    Quantum networks could be operating in the same infrastructure (e.g., fiber, free-space link) as classical optical communications, within which the connection is preliminarily established by distributing entangled-photon pairs. From the perspective of information service, the future quantum internet is a manner of configuring and scheduling relevant resources to satisfy people's requirements of accessing distinguishable quantum advantages, instead of the thin definitions of necessary hardware and software components with their services of entanglement distribution and so on. In this sense, all served user-cases, whether distributed sensing platforms or secure quantum communications, could be regarded as nodes connected to resources hub through quantum channels that are majorly comprised of optical fibers. Common resources the nodes demand are not limited to distributable photon pairs generated by entangled-photon sources (EPSs). For example, the high-precision quantum-signal detector is also important. As the final step of most quantum experiments, single-photon detecting performance is often the bottleneck for coincidence counting rates, purity examination, etc \cite{active-routing-QLAN-2013,zhang2008distribution}. Relevant research with single-photon detectors (SPDs) would be considerably advanced should they become available resources accessible by all users via fiber links. The fiber linkage itself is also a kind of available network resource. A flexible fiber-based network should allow users' platforms (e.g., EPS) to be common resources, which further provide corresponding services for the entire network. Therefore, we made an exploratory extension of the functional definition of local quantum networks: quantum resources are distributed to remote nodes via quantum channels for users' self-defined quantum applications. Under this definition, the quantum networks are no longer a series of stacked applications but represent infrastructures providing services in form of uniformly defined quantum resources. By analogy to the internet service provider (ISP) technology that grants individuals assess to the internet and affiliated services including domain name registration, web hosting, usenet service, colocation, etc., such a reconfigurable quantum network protocol can be defined as a quantum-version ISP, or, a QISP that serves a set of local individuals via entanglement-based connection. Any communication with quantum signals/entities as information carries could fall into its service purview. Apart from the protocol of QISP, some basic properties it has to fulfill are required:
    \begin{itemize}
        \item \textbf{Point-to-point entanglement connection.} Entangled photon pairs can be distributed to any two nodes.
        \item \textbf{Non-conflict concurrent service.} Parallel requests from multiple nodes are in the range of its service capacity.
        \item \textbf{Reconfigurability.} Reconfigurability means the adaptivity to heterogeneous user-cases that have sufficient freedom to configure their own quantum experiments. On the other hand, it means the capability to be easily modified and implanted in both hardware and software layers.
        \item \textbf{Software-defined network.} It must be in a software-defined network, allowing a series of software stacks seamlessly to integrate all hardware platforms and provide real-time dynamic scheduling.
    \end{itemize}

    With so many demanding features of this goal, it is so challenging to develop a truly layered, dynamically reconfigurable, and real-time schedulable quantum network. Recently some works on entanglement distribution in local quantum networks demonstrate enlightening implementation of relevant components. One kind of software-defined quantum network \cite{active-routing-QLAN-2013} is verified applicable in a small entanglement-distributed network with one EPS and eight users. In that demonstration, any two users can achieve entanglement by actively controlling fiber routers. Although the fidelity of distributed polarization-entangled photons achieved 99\%, the distances among nodes are too short to present convincing practicality of their results. Furthermore, the active control to establish entanglement is not sufficient to show features of a software-defined quantum network. Another implementation of layered local network presents pretty performance among intermediate-distance lab nodes \cite{RQLAN-2021}. While the feature of dynamic configuration proposed in it is just with respect to conceptual design but not practically scalable demonstration. Moreover, all the existing local quantum networks are controlled via primitive electronically automatic manipulation. Either the functional components or the network architectures of them are far from a reasonable implementation of QISP. 

    \begin{figure}[tb]
      \centering
      \includegraphics[width=\columnwidth]{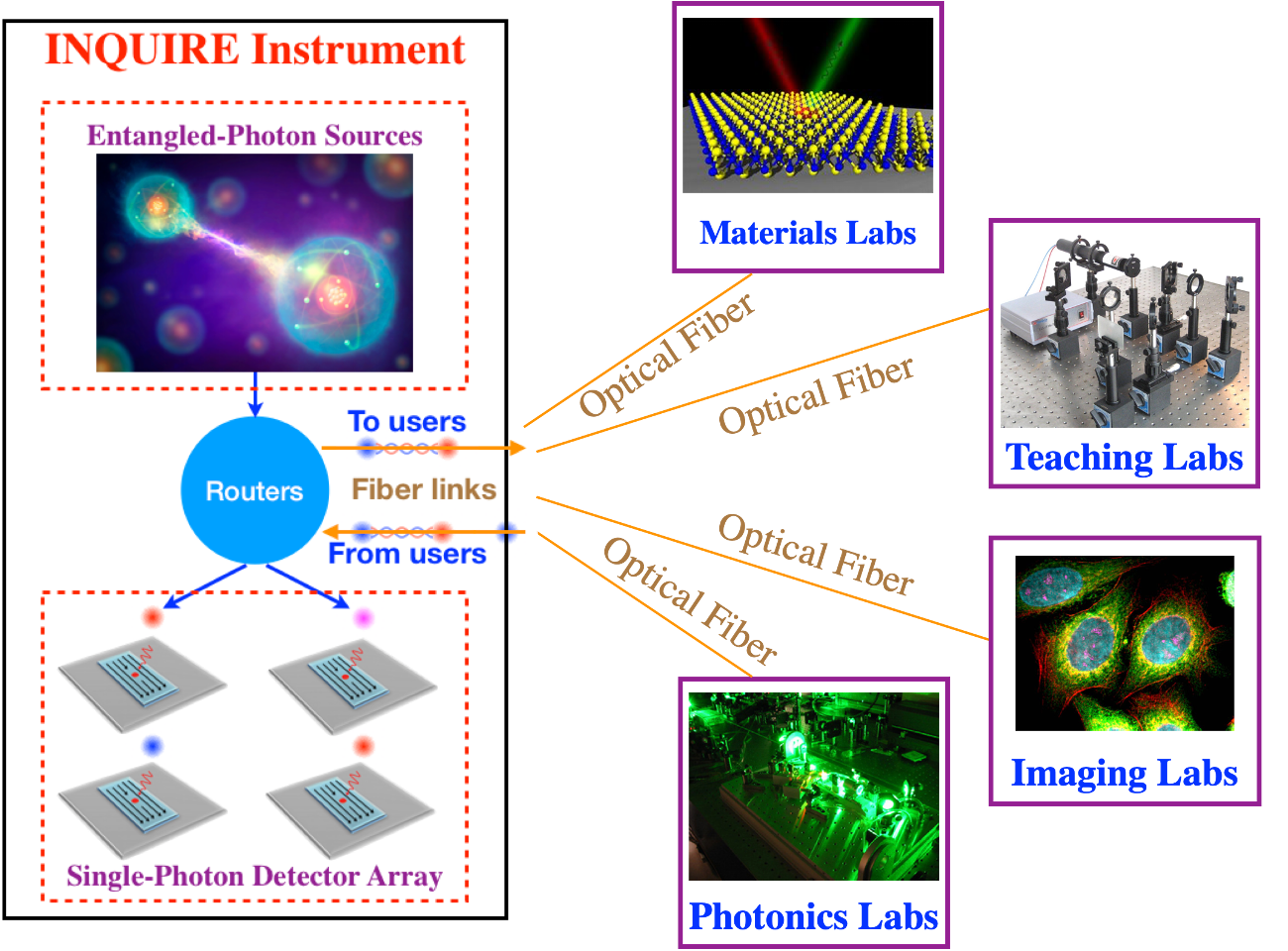}
      \caption{The deployed INQUIRE instrumental system serves versatile research and teaching objectives via fiber-based transmission of quantum signals. INQUIRE consists of high-quality EPS and SPD resources and quantum routers controlling photon streams between terminal labs and the hub.}
      \label{fig:profile}
    \end{figure}
    
    In this work, we adhere to a top-down vision to develop a sophisticated QISP prototype. A series of self-contained components and architectures are designed and implemented on the fiber-based quantum network at UA. Specifically, we construct our campus quantum network into the Interdisciplinary Quantum Information Research and Engineering (INQUIRE) testbed that re-integrates core instrumental resources and re-defines their specific functionalities in it. The core instruments are versatile EPSs and an eight-channel superconducting nanowire single-photon detector (SNSPD) array. They serve at most 16 lab nodes, each of which is able to access at most 5 EPS channels and 4 SNSPD channels, via dynamic configuration of optical switches that serve as quantum routers. While the developed highly modular software-related parts are decoupled from the testbed, serving as a universal Web-based operation software framework dubbed Quagent (Quantum Agent). In the B/S software architecture, Quagent allows for quantum resource accessing in the ``request-response'' internet service mode. Quagent has been practically deployed, continuously serving multiple quantum experimental platforms involved in the testbed, including color center spin systems, integrated nonlinear photonics, biophysical sensing, and fundamental quantum information research platforms. In this sense, the whole is a service-oriented Platform-as-a-Service (PaaS) system. All terminal labs’ platforms are off-the-shelf network components. Request processing and resource scheduling as well as real-time data acquisition are automatically completed by Quagent. Users can configure their own quantum applications and access real-time experimental results via the cloud. Quagent adapts to our testbed well even with some reconfiguration, e.g., modification of instruments alignment or connected interfaces, and it is implantable to other similar local quantum networks, only with modification of some soft-hardware interfaces. To characterize the key components --- the EPSs and verify the testbed's functionalities, we demonstrated multi-channel distributing and reconfigurable routing telecom-band time-energy entangled photons among different buildings with distances $ \sim 1$ km. Furthermore, we performed field tests of concurrent services for multiple users, where resource allocation/recovery and data acquisition are in instant dynamic response. The quantitative evaluation demonstrates the effectiveness, robustness, and scalability of the QISP prototype, but also proves to be helpful to optimize both hardware network structure and software scheduling strategy. Our implemented prototype also paves the way to the ``server-level'' scientific instrument research, instead of the ``standalone'' laboratory setting, in line with the goal of creating a widespread impact with quantum advantages in the upcoming quantum era. 

    The rest of this paper is structured as follows. In Sec. \ref{sec:2:resutls}, we firstly describe the hardware infrastructures and their re-constructing approaches on basis of the existing fiber-based network at the University of Arizona (UA), and then explain how the QISP is implemented under a reconfigurable framework. Field-test evaluation results are also included in this section. We summarize our work in Sec. \ref{sec:3:discussion} with analyses of current limitations and future work directions. In Sec. \ref{sec:4:methods}, we will see in more detail the Quagent's architecture and the entanglement source characterization methods.

\section{Results}\label{sec:2:resutls}

\subsection{Platform settings}
\begin{figure}[tb]
        \centering
        \subfigure[Layout of INQUIRE testbed at the UA main campus.]{
                \includegraphics[width=\columnwidth]{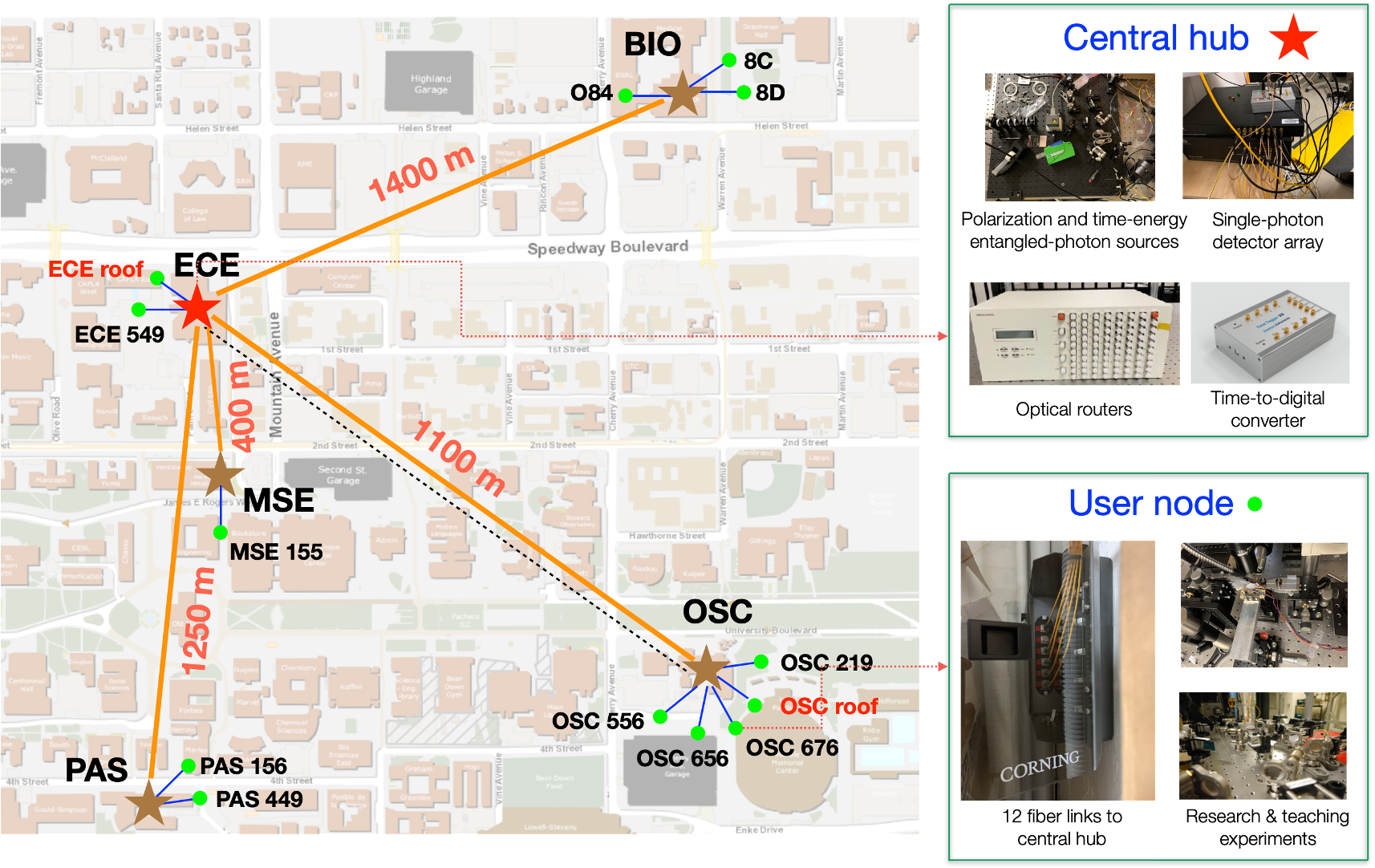}
        } 
        \subfigure[Configuration of the quantum routers for photon routing control.]{
            \includegraphics[width=\columnwidth]{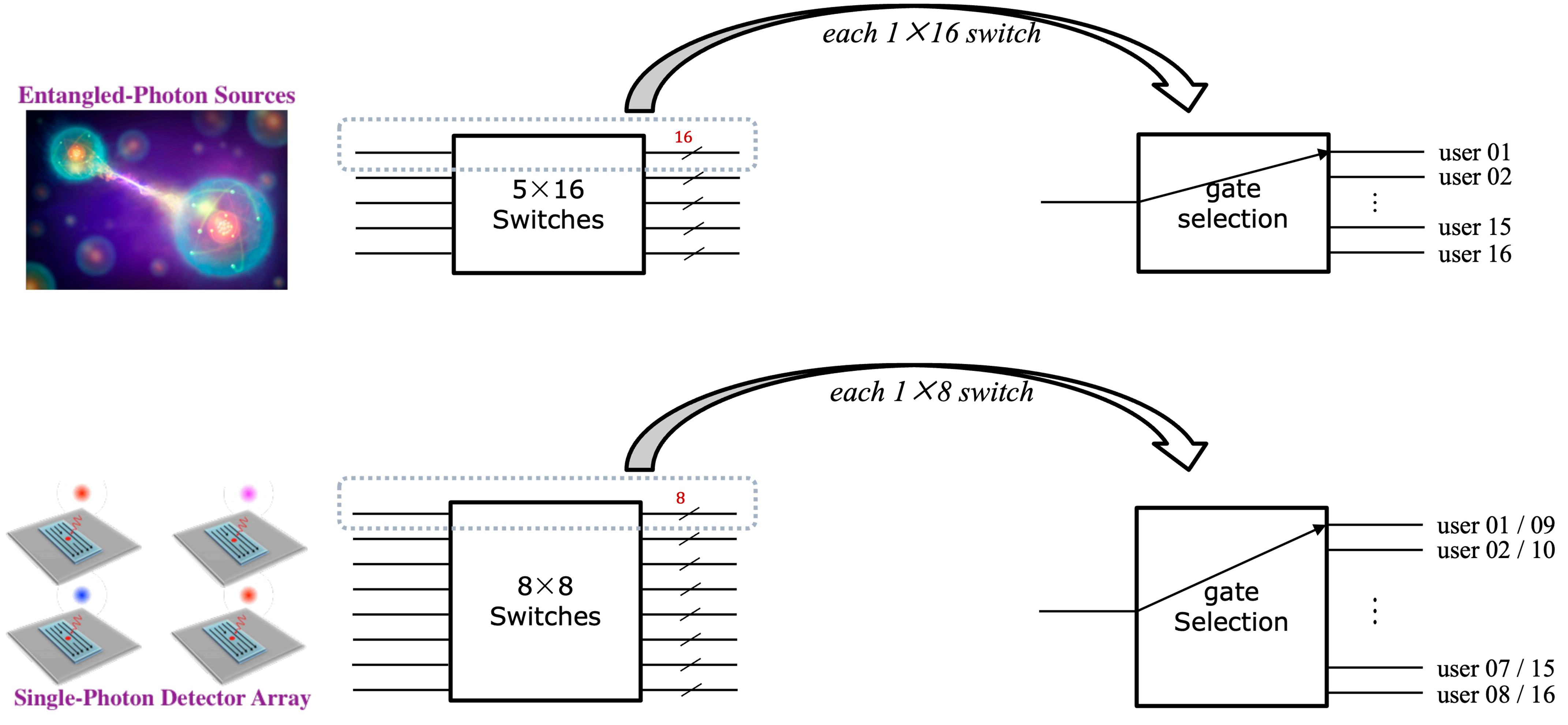}
        }
        \caption{Layout and configuration of INQUIRE.\quad(a) INQUIRE has a star topology comprised of 5 buildings as hubs and 13 labs as terminal nodes. The central hub (ECE building) is directly or indirectly connected to the 13 nodes located in the Materials Science and Engineering (MSE) building, the Physics-Atmospheric Sciences building (PAS), the College of Optical Science (OSC) building, and the BIO5 Institute (BIO) building.\quad(b) The 5x16 and 8x8 optical switches serve as bandwidth-unlimited quantum routers. Each user is configured with 5 EPS channels and 4 SPD ones. Each 1x16 switch connects one EPS channel to 16 users ($ \mathrm{user}_1\sim\mathrm{user}_{16} $); each 1x8 switch connects one SPD channel to 8 users ($ \mathrm{user}_1\sim\mathrm{user}_{8} $ or $ \mathrm{user}_9\sim\mathrm{user}_{16} $). For EPS resources, each user can access any one of 5 EPS channels but one EPS can be only occupied by one of 16 users at any time; for SPD resources it is similar.}
        \label{fig:layout-config}
    \end{figure}

    The INQUIRE instrumental system leverages the constructed specific fiber network at UA and integrates multi-access EPSs and SPD array. As an unprecedented architectural exploration of near-term local quantum networks, INQUIRE removes the barrier to generation and maintenance of quantum resources for a broad scientific audience, that is, EPSs and SPDs are assembled in an accessible platform that can communicate with served lab nodes or the so-called user-cases, over the campus fiber-based network, between buildings, colleges and scientific disciplines. The scale of our testbed is much larger than existing local quantum networks \cite{RQLAN-2021,garcia2021madrid}, and more importantly, it achieves the anticipation of versatile services. Correspondingly, INQUIRE is deployed in a two-level star-topology layout as shown in Fig. \ref{fig:layout-config} (a), comprised of commercial single-mode fiber patch cords with low loss at the telecom band, which is compatible with classical optical communication networks. Each link in fact represents a bunch of telecom-band fibers to be used for two-way communication. More than a dozen laboratories affiliated with five buildings are involved in the testbed, divided into ``hubs'' and ``terminal nodes''. The central hub resides in the Electrical and Computer Engineering (ECE) building, where all EPSs and the SPD array are located. The interaction between lab nodes and core resources is completed by two-way ``routers'' under dynamic scheduling. The overall layout coincides with our ``server-client'' design philosophy and proves to be of much scalability. Although all physical linkages compose a sparse star-topology graph, entanglement correlation could be established between any two of 13 terminal nodes, and thus it is a 13-node complete graph in the logical layer. 
    
    As the entanglement-based connection is the central functionality of INQUIRE, several state-of-the-art polarization-entangled and time-energy-entangled EPSs serve as key components to support heterogeneous applications, including teaching labs, quantum computing, quantum communication, biomedical sensing and imaging and so on. All entangled photons are generated from the spontaneous parametric down-conversion (SPDC) process of nonlinear crystals (type-II PPKTP bulk for discrete-variable (DV) polarization entanglement, type-0 PPLN for continuous-variable (CV) time-energy entanglement), under excitation of pulsed or continuous-wave (c.w.) lasers. On average, the generated DV and CV entangled photon fluxes are at rates of $10^6$ pairs/s and $10^{10}$ pairs/s, respectively. With configuration of telecom C-band (1530$\sim$1565 nm) coarse wavelength demultiplexers (CWDMs), the wavelength of entangled photons gets selected. An eight-channel SNSPD array (base temperature $ T < 2.8$ K, quantum efficiency $ \eta > 80\% $ at 1550 nm, 70 ns FWHM jitter) connected to an eight-channel time tagger (streaming time-to-digital converter, 80 ps FWHM jitter, 8.5 M count rate) provides the detecting functionality.

    INQUIRE's service capacity is defined by five different kinds of entanglement resources and an SNSPD's eight equivalent detecting channels. Correspondingly, two sets of optical fiber switches are connected between user nodes and ECE hub, thus fulfilling active control of photon streams from EPSs and to SPDs, respectively. The specific linkage configuration of switches is illustrated in Fig. \ref{fig:layout-config} (b). Considering the daily usage frequency of involved laboratories at UA, even without CWDMs to multiply entangled-photon channels, the service capacity is sufficient for served objects in this stage. At least, the current switches configuration scheme is able to support 16 terminal nodes, thus there is still 3-node allowance cause currently only 13 laboratories are involved in INQUIRE.

\subsection{QISP implementation and evaluation}
    The eventual goal is to construct INQUIRE testbed into a reconfigurable QISP prototype. Apart from those high-quality instrumental parameters, abstract and modular designs and state-of-the-art engineering techniques are pretty necessary to fulfill the required properties of QISP. 
    
    \begin{figure}[tb]
        \centering
        \includegraphics[width=\columnwidth]{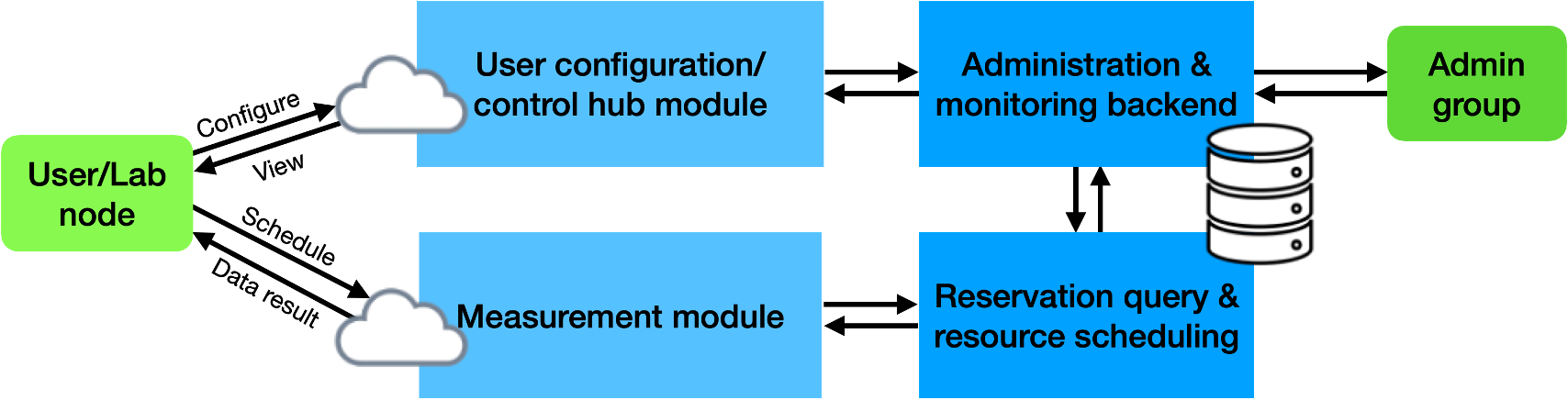}
        \caption{Brief implementation framework of Quagent.}
        \label{fig:quagent-strategy}
    \end{figure}

    Building on the well-defined quantum resources and operation components, we further take a ``request-response'' scheme to implement the service system in form of a Web-based software framework depicted in Fig. \ref{fig:quagent-strategy}. Since high-quality EPS and SPD resources are key components of most quantum experiments, they are described as Web-service resources accessed via uniform RESTful (Representational State Transfer) interfaces. The systematic Web-based control software integrates a set of reconfigurable modules and functionalities, including soft-hardware interfaces, resources scheduling, real-time data acquisition and visualization, network status monitoring, etc. It is a software-layer agent for local quantum networks and thus is named Quagent. Once a user is occupying some resources defined in Quagent, he can fulfill any relevant quantum applications he wants, such as quantum teleportation, photon correlation detecting, etc. Although these user-cases are heterogeneous, what the service hub provides is uniform quantum resources, and services are performed in a plug-and-play manner in the systematic B/S-architecture software. Therefore, it is a PaaS mode. Hardware and software infrastructures are transparent for users who only need to be concerned about what and how many quantum resources they want. Quagent will record all successful resource reservations and schedule resource allocation and recovery tasks on time. Furthermore, Quagent visualizes all detecting outcomes in real-time and makes them accessible like local measurement operations.

    \begin{figure}[tb]
        \centering
        \includegraphics[width=\columnwidth]{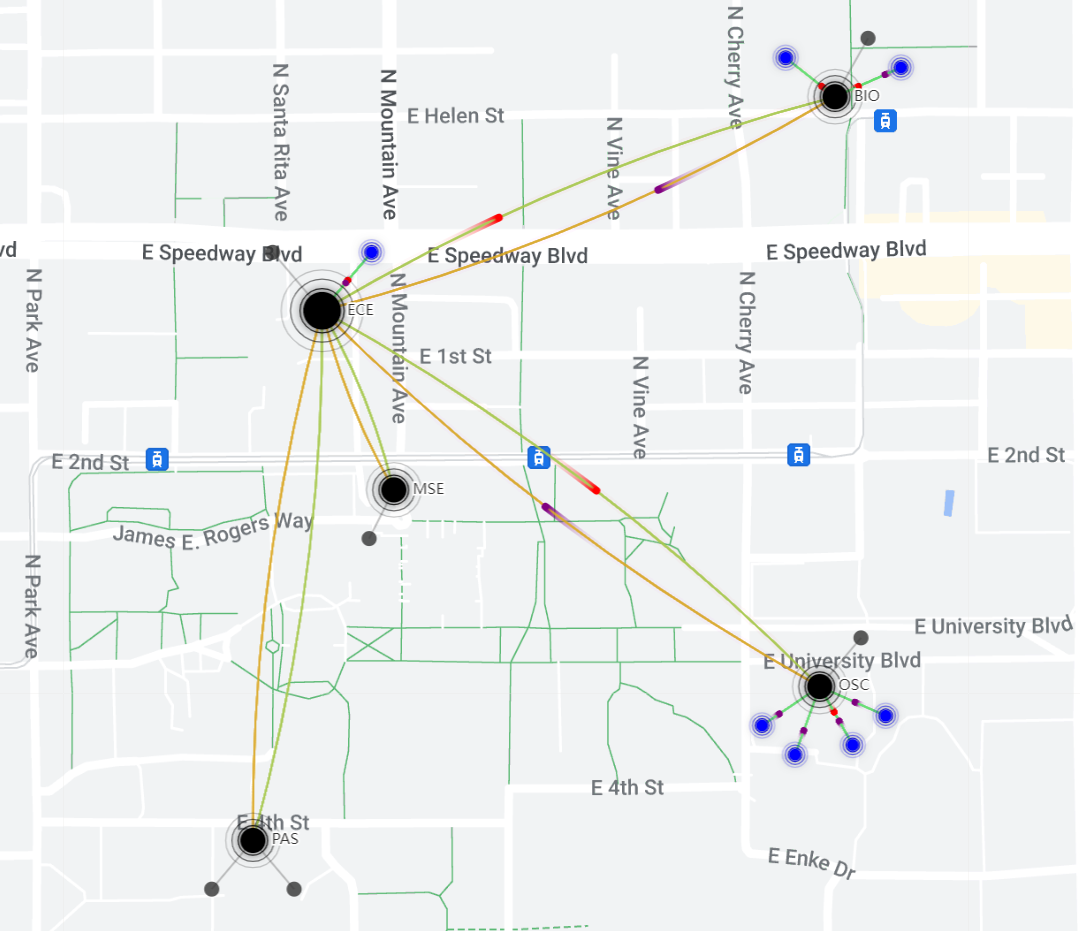}
        \caption{Real-time dynamic status of network testbed. Quagent is serving multiple users (lab nodes) simultaneously. Rippling blue nodes represent active users occupying EPS or SPD channel resources; grey nodes represent inactive users; large rippling black circles are hubs located in five buildings while ECE is the central hub. Tailing red circles represent entangled-photon streams; tailing purple circles represent single photons from users to be detected.}
        \label{fig:network-status}
    \end{figure}
    
    As aforementioned, the DV-based polarization-entangled EPS is widely adopted in related works, cause they are naturally compatible with solid-state qubit systems, especially in quantum computing. In addition to polarization-entangled EPSs, the integration of time-energy EPSs is a conspicuous characteristic and advantage of our platform. They together provide versatile quantum applications. In our kilometer-range network, time-energy entanglement is more practical for transmission due to its insensitivity to polarization-mode dispersion. Hence we focus on a fine-granularity performance characterization of a representative time-energy EPS of INQUIRE. By a field test of distribution and reconfigurable routing of multi-channel entangled photons in this testbed, the functionalities got well verified \cite{cui2021entanglement}. The entanglement is witnessed by the non-local dispersion cancellation phenomenon. The sharp temporal correlation spectrum is eventually obtained in scenarios of different dispersion scales and light wavelengths. Related experimental details are described in Sec. \ref{sec:4:methods}. It confirms both the high fidelity of distributed entanglement and the capability of engineering dispersion between multiple pairs of links. 
    
    To evaluate the scheduling capacity of Quagent, we performed field tests that multiple users request EPS and SPD channels in subsequent and then concurrent services take place. Results show that the processing capability of ECE hub with Quagent on photon routing and task scheduling is sufficient for maximal concurrent scenarios. The network status is real-timely monitored as Fig. \ref{fig:network-status} illustrates.

\section{Discussion}\label{sec:3:discussion}

    From the perspective of regarding quantum networks totally as QIS infrastructure, quantum networks' potentials get better exploited. That can be formalized to the QISP paradigm we have proposed. Our exploratory QISP implementation --- INQUIRE, is a reconfigurable fiber-based network testbed, demonstrating enormous potentials of near-term and next-generation quantum networks. With the integral operation via Quagent, our INQUIRE instrumental system proves to be a well-evaluated QISP prototype. It is a successful implementation and has been applied at UA for 13 lab nodes' on-demand services, under continuous optimization. This significant implemented QISP serves as a shared major facility that has fostered a series of interdisciplinary research and teaching missions in QIS.
    
    QISP is far beyond the scope of conventional research and application of quantum networks, at least at the local-area scale. For instance, in the quantum encryption field, INQUIRE platform is totally compatible with the Key-as-a-Service paradigm \cite{white2022quantum}. The INQUIRE testbed provides for a means to deliver unique quantum resources to multiple scientific disciplines and offers a first-of-its-kind platform for exploring the beginnings of quantum information sharing and processing. The open-sourced Quagent software operation framework that is composed of all software-related parts is highly integrated but decoupled with instrumental parts of INQUIRE, aiming at accelerating innovation of a practical quantum network design and reconfiguration. The advanced computer engineering techniques and highly modular architecture of Quagent allow for its reconfigurability and implantability to any other similar fiber-based networks. 

    Although we have not implemented a larger-scale, more-functionalities, higher-performance QISP due to restrictions of our hardware platforms, the satisfactory properties and guiding implementation rules in this work are favorable to deploy an industry-level QISP. On the other hand, hopefully these architectural insights are able to boost exploring innovative approaches to constructing and utilizing quantum networks.

    For future work, there are still many topics requiring to be explored for a truly next-generation QISP realization. At the component level, more requirements require to be taken into account as the community of served users scales up. For the functionality of entanglement distribution, CWDMs and corresponding software control modules need to be integrated into our testbed, to multiply EPS channels. In fact, CWDMs can be regarded as bandwidth-limited quantum routers in comparison with optical switches that are bandwidth-unlimited. For the functionality of single-photon detecting, measurement synchronization will become an important issue when measurements of every part of photon pairs are done in different locations or different time-to-digital converters. Even our current centralized resources alignment subtly avoids this problem, it is inevitable once our testbed becomes larger and more than one central hub is involved. The scheme of GPS-assistant synchronization proves to be effective in recent entanglement distribution experiments \cite{RQLAN-2021}, but there is still much room to improve the synchronization precision. At the architecture level, the adaptivity to physically heterogeneous quantum networks is necessary to be satisfied. That is, future quantum networks have to beyond the only fiber-based communication approach. Now we have  established free-space communicative links from ECE to OSC and are attempting to integrate them into INQUIRE. In the longer term, it will also be developed and tested that the satellite-assistant entanglement distribution between the Tucson testbed and Boston testbed, in which the latter is committed to building novel quantum repeaters of the NSF Center for Quantum Networks (CQN). At the application level, some innovative quantum applications benefiting from our INQUIRE testbed are worth exploring or verifying, such as the distributed CV sensing with multi-part communication served by the central hub. Particularly, when a system becomes larger and more complicated, some potential issues might concomitantly occur, which proposed higher requirements of optimization or refactoring for specific QISP architectures. On the other hand, reasonable evaluation metrics of QISP systems need to be defined and standardized, requiring efforts from the whole QIS community.

\section{Methods}\label{sec:4:methods}

\subsection{Multi-channel entanglement distribution and routing}

    The field functionality verification and parameters characterization are performed in parts of INQUIRE testbed, and the demonstrative multi-channel distribution and routing scheme could be directly scaled to all INQUIRE's components. Different from existing entanglement verification approaches, we witness the distributed entanglement by non-local dispersion cancellation effect of time-energy entangled photons \cite{cui2021entanglement}. 

    As one promising candidate for long-distance fiber-based transmission, time-energy entangled photon pairs at telecom waveband suffer from low propagation loss in standard optical fiber and insensitivity to polarization-mode dispersion compared with polarization-entangled photons. The slight dispersion can be further suppressed by means of non-local dispersion cancellation effect \cite{franson1992nonlocal}, in which the dispersion experienced by one photon can be canceled by the dispersion experienced by the other photon. This effect is underpinned by non-classical time-energy correlations between a pair of photons and it is independent of the distance between them.

    \begin{figure}[tb]
        \centering
        \subfigure[Experimental setup]{
            \includegraphics[width=\columnwidth]{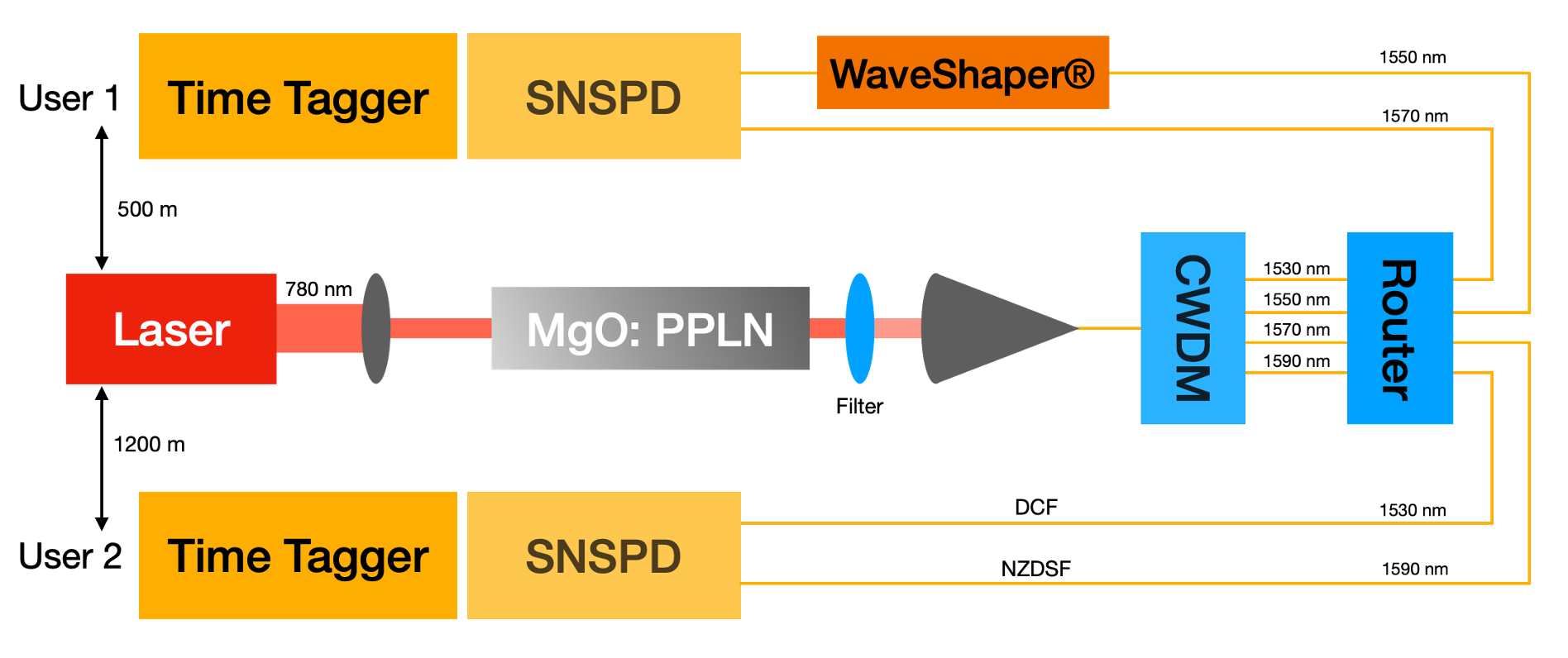}
        }\\
        \subfigure[Biphoton time-correlation measurement result]{
            \includegraphics[width=\columnwidth]{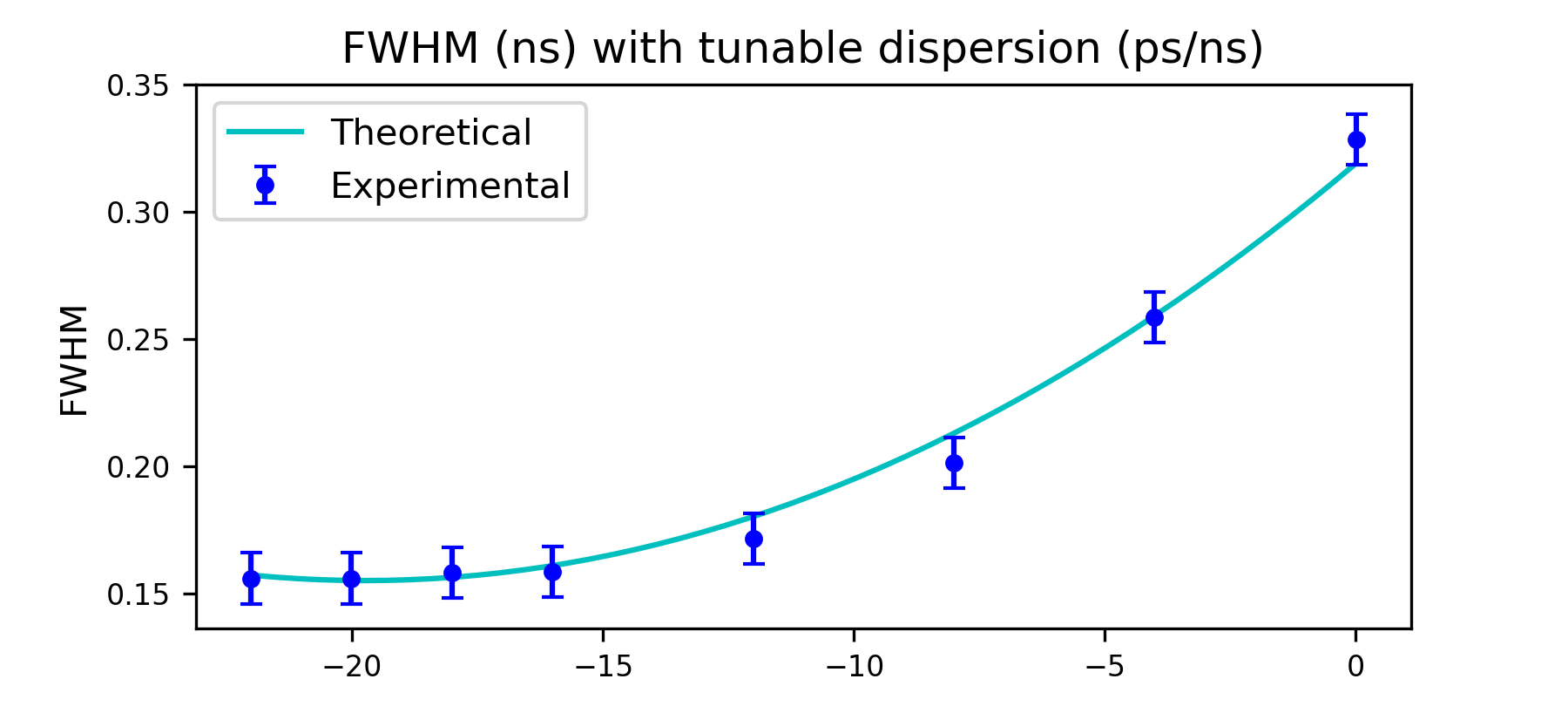}
        }
        \caption{Scheme and result of time-energy entanglement characterization. (a) Setup for testing multi-channel photon routing and the non-local dispersion cancellation. DCF: dispersion compensating fiber; NZDSF: non-zero dispersion-shifted fiber. (b) FWHMs of measured time correlations between the 1570-nm photons and the dispersion-compensated 1550-nm photons traveling from the time-energy entanglement source to User 1. The dispersion compensation applied on 1550-nm photon varies from 0 to $ -22 $ ps/nm where the minus sign means compensating the fiber's normal dispersion. The error bars represent the uncertainty of Gaussian fitting.}
        \label{fig:charac}
    \end{figure}
    
    In our field test, we generate broadband time-energy entangled photons by the type-0 SPDC process of a MgO-doped PPLN bulk crystal pumped by a c.w. 780-nm laser. The entangled photon pairs are separated by a C-band CWDM into 8 channels each with a 16-nm bandwidth. Two pairs of channels centralized around 1550/1570 nm and 1530/1590 nm are used in tests of entanglement distribution and routing among three buildings with kilometer-range distances. Correlated-detecting outcomes of photons are outputted by the high-precision SNSPD and time tagger as described above. All these procedures are automatically and remotely controlled via our software modules.
    
    To witness the biphoton entanglement phenomenon, it is only needed to compensate anomalous dispersion on one fiber channel. Take the case of biphoton pairs sent to User 1 as shown in Fig. \ref{fig:charac} (a), we insert a WaveShaper in to add tunable anomalous dispersion in a 16-nm band at around 1550 nm. The recorded biphoton temporal correlation histograms are fitted with a Gaussian function. Fig. \ref{fig:charac} (b) illustrates how the fitted full width at half maximum (FWHM) of the correlation spectrum varies with the tunable dispersion scale. In our setting, the single-mode fiber has an estimated dispersion of 9.7 ps/nm. When the WaveShaper overcompensates the 1550-nm photon flux at the dispersion scale of -19.4 ps/nm, i.e., the negative twice the fiber dispersion, the almost narrowest FWHM is observed by User 1. The remaining uncertainty is due to our detector jitter limited, showing agreement with the theoretical prediction.

\subsection{Software framework and deployment}

\begin{figure}[tb]
        \centering
            \includegraphics[width=\columnwidth]{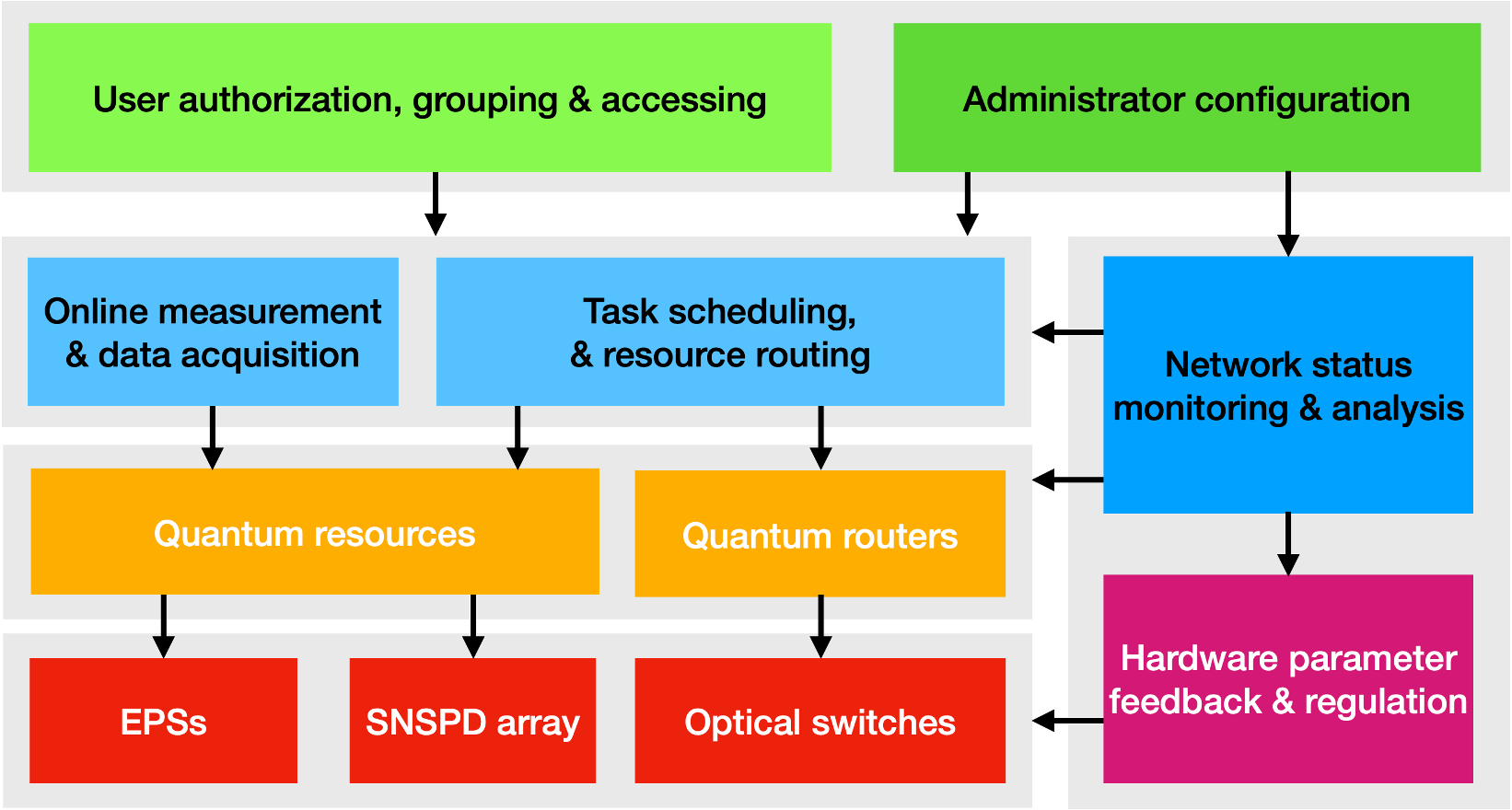}
        \caption{Architecture and functional modules of Quagent. From bottom to up, they are the hardware layer, resource layer, scheduling layer and user administration layer. Additional modules (right part of the figure) aim at regulating and monitoring the whole platform.}
        \label{fig:quagent-modules}
    \end{figure}

    Quagent takes use of the RESTful programming standard to define instruments as abstracted Web resources. In appearance, as depicted in Fig. \ref{fig:quagent-strategy}, authorized users communicate with Quagent server via HTTP request-response protocol, while each successful Web communication represents heralded quantum resources allocation and non-classical information transmission. These user-friendly procedures are supported by a series of backend modules of Quagent, as illustrated in Fig. \ref{fig:quagent-modules}. All implemented modules and interfaces are integrated in a hierarchical way and they are decoupled with each other, leading to the anticipated reconfigurability. For instance, since the devices (EPS, SNSPD, etc.) are defined as abstracted instrumental resources to be managed in Quagent, Quagent is able to be easily implanted to other hardware platforms by just modifying the implementation of the specific interfaces. Meanwhile, the affiliated user-friendly frontend interactive pages make it totally convenient for users to perform on-demand quantum applications. To achieve high-performance persistent service, advanced databases and robust Web servers are configured to deploy Quagent on the testbed, which runs on a specific computer server residing in ECE hub.
    

\section{Open-source Quagent deliverable}

    Our Quagent project is available on \href{https://github.com/Youngcius/quagent}{GitHub} with rich demonstrative tutorials and manuals. Detailed software technical strategies and hardware devices description are all accessible on it. Quagent uses Apache-2.0 open-source license. Although it is a PaaS framework closely dependent on corresponding hardware platforms rather than a pure-software application, we hope researchers in related fields could rapidly prototype a similar or extended QISP system by means of the design modes, adaptive modules, deployment manners, hard-software interfaces and documentations of Quagent, while on their own local-area quantum networks.

\section{Acknowledgement}
    This work is supported by an NSF Major Research Instrumentation project (No.~ECCS-1828132) and the Center for Quantum Networks (No.~EEC-1941583), one of the fourth-generation NSF Engineering Research Centers (ERC) established in 2020. We gratefully acknowledge the in-kind support by Corning Inc. in 2019 of the optical fibers deployed in INQUIRE.  We thank favorable feedbacks from the CQN community.


\bibliographystyle{IEEEtran}
\bibliography{reference}



\end{document}